\documentclass[12pt]{article}

\usepackage{amsmath}
\usepackage{amssymb}
\newcommand{\cC}{{\cal C}}
\newcommand{\cF}{{\cal F}}
\newcommand{\cH}{{\cal H}}
\newcommand{\N}{{\cal N}}
\newcommand{\cN}{{\cal N}}
\newcommand{\cO}{{\cal O}}
\newcommand{\cS}{{\cal S}}
\newcommand{\IZ}{{\mathbb Z}}
\title{Families of $N=2$ field theories}
\begin{document}
\maketitle
The main actors of this review are four-dimensional field theories with $\cN=2$ supersymmetry. There are three well-understood  ways to build large classes of $\cN=2$ field theories\footnote{It is also possible to define four-dimensional $\cN=2$ field theories from a circle compactification of a $\cN=1$ 5d SCFT, or a torus compactification of a 
six-dimensional $(1,0)$ SCFT. We are not aware of four-dimensional $\cN=2$ field theories which can only be built that way.}: \begin{itemize}
\item Standard four-dimensional Lagrangian formulation  
\item Twisted compactification of a six-dimensional $(2,0)$ SCFT (``class $\cS$'')
\item Field theory limit of string theory on a Calabi-Yau singularity (``geometric engineering'')
\end{itemize}
These three classes of constructions have large overlaps. 
Most four-dimensional Lagrangians can be engineered in the class $\cS$, 
and all can be engineered through some Calabi Yau geometry. Class $\cS$ theories can be further lifted to Calabi Yau compactifications involving a curve of ADE singularities. Conversely, only a minority of $N=2$ field theories admits a direct four-dimensional Lagrangian description. 

Different UV realizations of the same theory may be better suited to answer specific questions. The six-dimensional or string-theoretic descriptions of a theory can be very powerful for computing properties which are somewhat protected by supersymmetry. 
On the other hand, some properties, symmetries and probes of a four-dimensional field theory may simply not be inherited from a specific UV definition of the theory. Simple four-dimensional field theory constructions may be hard to lift to six dimensional field theory, and even harder to embed in string theory, where every modification of the theory must involve a dynamical configuration of supergravity fields and D-branes which solves the equations of motion. A reader of this special volume will have several occasions to appreciate the power of these alternative approaches.  

This chapter of the review is intended essentially as a reading guide. We refer the reader to the original references and many excellent reviews available to learn the basic properties of $\cN=2$ field theories. We do not feel we can improve significantly on that available material. We will try to present a global overview of more recent developments.

\section{Lagrangian theories}
The requirements of $\cN=2$ supersymmetry and renormalizability impose very strong constraints 
on the possible couplings in a Lagrangian \cite{WB}. We will assume the reader has some familiarity with the construction of 
$\cN=1$ supersymmetric Lagrangians. Already for an $\cN=1$ theory most of the freedom would lies 
in the choice of superpotential for the theory. Requiring the presence of $\cN=2$ supersymmetry fixes the form of the superpotential. As a result, the possible $\cN=2$ renormalizable Lagrangians are labelled by a choice of gauge group and of the representations the matter fields sit in. 

The gauge fields belong to vectormultiplets, which decompose into a $\cN=1$ gauge multiplet and an adjoint chiral multiplet $\phi$. The simplest Lagrangian $\cN=2$ theories are pure gauge theories. The chiral multiplet $\phi$ has no superpotential, and the only free parameter in the Lagrangian is the complexified gauge coupling 
\begin{equation}
\tau = \frac{4 \pi i}{g_{YM^2} }+ \frac{\theta}{2 \pi}
\end{equation}
The beta functions for the gauge couplings are one-loop exact, and non-Abelian theories are asymptotically free. Abelian gauge groups coupled to matter, 
on the other hand, are IR free and have a Landau pole. The can only appear in effective theories. 

The vectormultiplet kinetic terms can be written as an integral over chiral $\cN=2$ superspace: we can assemble the vectormultiplet into a chiral superfield 
\begin{equation}
\Phi(\theta) = \phi + \cdots
\end{equation}
depending on two sets of chiral superspace variables $\theta^1_\alpha$ and $\theta^2_\alpha$. 
Then the standard kinetic terms take the form
\begin{equation} 
{\cal L}_{\mathrm{kinetic}} = \tau \int d^4 \theta \mathrm{Tr} \Phi^2 + \mathrm{c.c.}
\end{equation}

A more general choice of kinetic term can be described by a local gauge-invariant holomorphic pre-potential ${\cal F}(\Phi)$, as 
\begin{equation} 
{\cal L}_{\mathrm{kinetic}} = \tau \int d^4 \theta {\cal F}(\Phi) + \mathrm{c.c.}
\end{equation}
This kind of expression can capture, say, the two-derivative part of a low-energy 
effective action. 

Matter fields can be added in the form of hypermultiplets, which in an $\cN=1$ language can be decomposed to a set of chiral multiplets $q^a$ sitting in a pseudo-real representation of the overall symmetry group, which is the product of the gauge group and possible flavor groups. 
A renormalizable $\cN=2$ Lagrangian can be written in $\cN=1$ superspace in a standard way, with a superpotential 
\begin{equation}
W_{\cN=1} = \mathrm{Tr}\phi \left( q^a t_{ab} q^b \right) + \mathrm{Tr}M \left( q^a t^f_{ab} q^b \right)
\end{equation}

Here $t_{ab}$ are the gauge symmetry generators (they are symmetric, as the representation is pseudoreal) and $t^f_{ab}$ are the flavor symmetry generators.
We will use the notation ``hypermultiplets in representation $R$'' to indicate a set of chiral fields in the representation $R \otimes \bar R$. 
A set of chiral multiplets in a pseudoreal representation $R$ will be denoted as a ``half-hypermultiplet in representation $R$''.
In some cases, discrete anomalies prevent half-hypermultiplets from appearing alone. 

The complex mass parameters $M$ live in the adjoint of the flavor group. They must be normal, $[M, M^\dagger]=0$, and can be thought as elements of the Cartan sub algebra of the flavor group. 

Thus the only parameters of standard UV complete Lagrangian $\cN=2$ theories are the gauge couplings and the complex mass parameters. \footnote{If Abelian gauge groups are present, one could turn on an FI parameter, which breaks explicitly $SU(2)_R$. As Abelian groups coupled to matter have Landau poles, and if an FI parameter is absent in the UV it cannot appear in the IR, they will rarely play a role in this review. A notable exception is the theory of BPS vortices, which can only occur in the presence of an FI parameter}

The matter representation is limited by the requirement of asymptotic freedom (or conformality). The beta functions for the gauge couplings are 
one-loop exact, and receive a positive contribution from every matter field. The limitation of asymptotic freedom allows a systematic classification of all possible Lagrangian $\cN=2$ gauge theories. The full classification and a very nice set of references can be found at \cite{Bhardwaj:2013qia}.

The simplest $\cN=2$ Lagrangian theories with matter are $SU(N)$ gauge theories coupled to fundamental hypermultiplets, i.e. $\cN=2$ SQCD, or to a single adjoint hypermultiplet, i.e. $\cN=4$ SYM (which is denoted as $\cN=2^*$ when the complex mass for the adjoint hypermultiplet is turned on). 
The beta function for the latter theory vanishes, and the gauge coupling is exactly marginal. 
For the former theory, the beta function vanishes if the number of flavors is twice the number of 
colors, i.e. $N_f = 2 N$. 

A larger class of examples are quiver gauge theories, built from $\prod_a SU(N_a)$ gauge theories
coupled to fundamental and bi-fundamental hypermultiplets. 
The constraint on the number of flavors implies that twice the rank of the gauge group at any given node must be bigger than the sum of the ranks at adjacent nodes. This is only possible if the quiver takes the form of a Dynkin diagram or, in the absence of fundamental matter, an affine Dynkin diagram. These theories will have two sets of flavor symmetries. Each bifundamental hypermultiplet (and symmetric or anti-symmetric hypers) is rotated by a $U(1)$ flavor symmetry. Each group of $M_i$ fundamentals at the $i$-th node is rotated by an $U(M_i)$ flavor group. 

Many more possibilities exist if we add matter in other representations, and look at more general choices of gauge groups. A possibility which will be important later is to consider $SU(2)^n$ gauge theories 
coupled to fundamental, bi-fundamental and (half)-trifundamental hypermultiplets. Tri-fundamental 
hypermultiplets are only allowed by renormalizability for three $SU(2)$ gauge groups, 
and their existence allows one to build intricate $SU(2)^n$ Lagrangian theories labelled by an arbitrary trivalent graph. 
This is not possible for other gauge groups. 

The low-energy dynamics of $\cN=2$ gauge theories is very rich, and mostly hidden in a UV Lagrangian formulation. Many interesting quantities, even protected by supersymmetry and holomorphicity, 
receive crucial perturbative and non-perturbative corrections.
Initially, the low energy dynamics was understood on a case-by-case basis from a careful analysis of the holomorphicity properties of the $\cN=2$ supersymmetric low-energy effective Lagrangian, starting from simple $SU(2)$ gauge theories \cite{Seiberg:1994rs,Seiberg:1994aj}. 

More systematically, many interesting results for very general quiver gauge theories can be computed by localization methods, with the help of the so-called $\Omega$-deformation of $\cN=2$ gauge theories. We refer to \cite{Nekrasov:2012xe} for a general analysis and references. Ultimately, as the our mathematical understanding of localization and of the geometry of instanton moduli spaces improves, we may hope to extend such calculations to all Lagrangian 
$\cN=2$ theories. 

An important alternative approach is based on string theory dualities. Many $\cN=2$ Lagrangian field theories 
can be engineered by brane systems and then mapped through dualities to configurations in M-theory \cite{Witten:1997sc} (but not for $E$ and $\hat E$-type quivers) and IIB geometric engineering (all quivers) \cite{Katz:1997eq}. 
The power of this approach lies in non-renormalization theorems which allow many protected quantities to be computed classically in the M-theory or IIB descriptions. 

These constructions can be thought as providing maps which embeds (most) of the $\cN=2$ Lagrangian theories into larger classes of $\cN=2$ quantum field theories, which are constructed through M-theory or IIB setups which equip each theory with a simple geometric description of its low-energy dynamics, but possibly not a straightforward four-dimensional field-theoretic UV description. These are the other two classes of $\cN=2$ theories mentioned int he introduction. 

Before exploring these classes of theories, it is useful to discuss some general properties of $\cN=2$ theories, abstracting from 
the possible existence of a Lagrangian description. 

\section{General properties of $\cN=2$ field theories}
Most of the facts collected in this section are easily demonstrated for Lagrangian field theories, but 
appear to be true for all $\cN=2$ UV complete quantum field theories.  It is likely that they could be established in 
full generality by an accurate analysis of the $\cN=2$ tensor and conserved current supermultiplets. 

The first general property is the existence of an $SU(2)_R$ R-symmetry. 
The $\cN=2$ SUSY algebra is compatible with an $SU(2)_R \times U(1)_r$ R-symmetry group.
Both factors appear as part of the $\cN=2$ superconformal group, and thus are always symmetries of $\cN=2$ SCFTs. The $U(1)_r$ is broken/anomalous 
for all asymptotically free or mass-deformed theories, as the breaking of the conformal and $U(1)_r$ symmetries are tied together by supersymmetry. 

In a Lagrangian theory, the hypermultiplet scalars sit in a doublet of $SU(2)_R$. In appropriate conventions, the top component of the doublets are $\cN=1$ chiral fields, the bottom are $\cN=1$ anti-chiral fields. The vectormultiplet scalars, on the other hand, are charged under $U(1)_r$. 
These fields are special examples of two important classes of protected operators: Coulomb branch operators and 
Higgs branch operators. These two classes of operators control both the parameter spaces of deformations and moduli spaces of vacua preserving $\cN=2$ 
supersymmetry. The geometry of these spaces is rich and plays a central role throughout this volume.  

Coulomb branch operators are operators annihilated by all anti-chiral supercharges: they are chiral operators for every $\cN=1$ sub algebra of the theory.
They never belong to non-trivial $SU(2)_R$ representations. In a SCFT they carry an $U(1)_r$ charge proportional to their scaling dimension. 
The Coulomb branch operators in a Lagrangian theory are holomorphic 
gauge-invariant polynomials of the vector multiplet scalar fields.  
For example, if we have some $SU(N)$ gauge fields with scalar super partner $\phi$, the traces $\mathrm{Tr} \phi^n$
are all Coulomb branch operators. 

A general $\cN=2$ may include many more Coulomb branch operators, not associated to weakly-coupled gauge fields. As long as their 
scaling dimension is smaller or equal to $2$, they will be associated to more general deformation parameters $c_i$ of the theory, written in chiral superspace as
\begin{equation}
\delta c_i \int d^4 \theta \cO^i
\end{equation}
involving the appropriate super-partner $Q^4 \cO^i$ of the Coulomb branch operators $\cO^i$.

The second class of deformations, complex masses, is also tied to vectormultiplets, but 
rather than being a coupling in a vectormultiplet Lagrangian, they are vevs of a background, non-dynamical vectormultiplet scalar fields. 
More precisely, if the theory has a continuous flavor symmetry, with Lie algebra $\mathfrak{g}$, we can couple it to a non-dynamical background vectormultiplet valued in $\mathfrak{g}$. A complex mass deformation is introduced by turning on 
 a vev $M$ for the complex scalar in the background vectormultiplet, such that $[M,M^\dagger]=0$. Up to a flavor symmetry transformation, we can take $M$ to be valued in the complexified Cartan sub-algebra of $\mathfrak{g}$.
 
At the level of the Lagrangian, the leading order effect of a complex mass is to add a coupling to a super partner $\mu^a_{++}$ of a conserved flavor current $J^a$
\begin{equation}
\delta M_a d^2 \theta^+ \mu^a_{++} + c.c.
\end{equation}
The $+$ refers to one component of a $SU(2)_R$ doublet index. 
Indeed, conserved currents sit in a special supermultiplet which includes an $SU(2)_R$ triplet of moment map operators $\mu^a_{AB}$. 

Moment map operators are the typical example of a Higgs branch operator: operators which sit in non-trivial $SU(2)_R$ representations of spin $n/2$, $\cO_{A_1 A_2 \cdots A_n}$,
and satisfy a shortening condition \cite{Dolan:2002zh}.
\begin{equation}
Q^\alpha_{(A_0}\cO_{A_1 A_2 \cdots A_n)} =0 \qquad \bar Q^{\dot \alpha}_{(A_0}\cO_{A_1 A_2 \cdots A_n)} =0
\end{equation} 
They never carry an $U(1)_r$ charge. 

\subsection{Parameter spaces of vacua and S-dualities}
As the beta functions of gauge couplings are one-loop exact, it is easy to construct conformal invariant Lagrangian $\cN=2$ field theories by tuning the total amount of matter appropriately. These Lagrangian theories will thus have a parameter space of exactly marginal deformations parameterized by the 
complexified gauge couplings. Although many isolated, strongly-coupled $\cN=2$ SCFTs exist, there are also large classes of non-Lagrangian $\cN=2$ 
SCFTs with spaces of exactly marginal deformations. Many examples can be defined by coupling standard non-Abelian gauge fields to 
the flavor symmetry currents of non-Lagrangian isolated $\cN=2$ SCFTs in such a way that the gauge coupling beta function vanishes. 

The space of exactly marginal deformations of an $\cN=2$ SCFT is a complex manifold, and several protected quantities 
are locally holomorphic functions on the space of deformations. \footnote{We would like to point out that a general analysis 
of the geometric properties of the space of exactly marginal deformations of $\cN=2$ SCFTs seems to be missing from the literature}.
Thus a side payoff of exact calculations, done by localization or M-theory/IIB engineering, is a characterization of the complex manifold of 
marginal couplings. 

The results, even for Lagrangian theories, are rather counter-intuitive. Naively, the space of couplings for a Lagrangian theory 
should consist of a product of several copies of the upper half plane, each parameterized by a complexified gauge coupling $\tau_a$. 
More precisely, as the gauge theory is invariant under $\tau \to \tau+1$, one can parameterize the space by the 
instanton factors $q_a = \exp 2 \pi i \tau_a$. (This is a good choice in asymptotically free theories as well, where $q_a$ becomes a dimensionful coupling. )

At weak coupling, protected quantities can be expanded in power series in the $q_a$, typically convergent in the naive physical range $|q_a|<1$. 
Surprisingly, with the exception of $\cN=4$ SYM, the geometry of parameter space is strongly modified at strong coupling, and $|q_a|=1$ is not a boundary anymore: the theory can be analytically continued beyond $|q_a|=1$ into a complicated moduli space. 

In all known examples, as soon as we move far enough from the original weakly coupled region new 
dual descriptions of the theory emerge, possibly involving radically different degrees of freedom. 
Typically, the new descriptions involve weakly coupled gauge fields interacting with intrinsically 
strongly-coupled matter theories described by isolated $\cN=2$ SCFTs. Only in some cases
we find again weakly-coupled Lagrangian theories. 

The generic moniker for this type of situation, where seemingly different theories are related by analytic continuation in 
the space of gauge couplings, is S-duality. The canonical example of S-duality occurs in $\cN=4$ 
SYM \cite{Montonen:1977sn,Sen:1994fa}: as one approaches the $|q|=1$ boundary, new dual descriptions emerge involving 
magnetic monopoles or dyons which reassemble themselves into weakly coupled $\cN=4$  gauge fields 
with the same gauge group of the original theory, or its Langlands dual group.
In that case, every description covers the whole parameter space, and the couplings are related by $SL(2,Z)$ 
transformations of the form 
\begin{equation}
\tau' = \frac{a \tau + b}{c \tau + d}
\end{equation}
which map the upper half plane back to itself.

The $N_f = 2 N$ SQCD already offers a more general situation: the full parameter space can be described 
by allowing $q$ to reach arbitrary values. At very large $q$ we have a dual $N_f = 2 N$ SQCD description, 
with coupling $q' = 1/q$. At $q \sim 1$ we have a non-Lagrangian dual description, 
with an $SU(2)$ weakly coupled gauge field of coupling $q'' \sim 1-q$, 
coupled to a fundamental hyper and to an isolated SCFT with $SU(2) \times SU(N_f)$ flavor symmetry \cite{Argyres:2007cn,Gaiotto:2009we}. 
Similar statements hold more general Lagrangian theories. The possible S-dual descriptions of Lagrangian theories 
which can be mapped to class $\cS$ are well understood. Other examples, such as the quiver theories in the shape of an E-type Dynkin diagram, do not appear to have been explored systematically. 

There is a neat class of examples of theories with the property that all S-dual descriptions are Lagrangian. This will be our introduction into the class $\cS$ theories. The starting point is the observation that for $SU(2)$ $N_f=4$ SQCD all three S-duality frames, 
around $q=0$, $q=\infty$, $q=1$, are described by a $SU(2)$ $N_f=4$ SQCD Lagrangian. As the $SU(2)$ 
fundamental representation is pseudoreal, the flavor group is really $SO(8)$, and the three S-dual descriptions are 
related by a triality operation of the flavor group: if the quarks in the $q \sim0$ description sit in the vector representation $8_v$, 
the quarks in the $q \sim \infty$ description sit in the chiral spinor representation $8_s$ and the quarks in the $q \sim 1$ description sit in the anti-chiral spinor representation $8_c$.

This beautiful result, originally found in \cite{Seiberg:1994aj}, has far reaching consequences. 
As we mentioned before, the existence of a trifundamental half-hypermultiplet for three $SU(2)$ 
groups allows the construction of a large class of $SU(2)^k$ Lagrangian field theories, 
with $SU(2)$ gauge groups only \cite{Gaiotto:2009we}. Each gauge group can be coupled to at most two trifundamental blocks, and will be conformal if coupled exactly to two. Thus we can associate such a superconformal field theory to each trivalent graph, with an $SU(2)$ gauge group for every internal edge and an $SU(2)$ flavor group for each external 
edge.

Two trinions coupled to the same $SU(2)$ essentially consist of four fundamental flavors for that group. 
A group coupled to two legs of the same trinion, instead, looks like an $SU(2)$ $\cN=4$. 
Thus if we start from a frame where all couplings are weak, and make a single coupling strong, we can apply one of the two basic S-duality operations to that group, and reach a new S-dual description of the whole theory. The new description is again an $SU(2)^k$ theory, but it is associated to a possibly different trivalent graph. Ultimately, all the theories associated to graphs with $n$ external edges and $g$ loops must 
belong to the same moduli space of exactly marginal deformations, represent distinct S-dual descriptions of the same underlying SCFT
labelled by $n$ and $g$. 

This will be the most basic example of class $\cS$ theories. Furthermore, these are the only S-duality frames for these theory. 
The parameter space of gauge couplings of these theories will be identified to the moduli space of complex structures for a Riemann surface of genus $g$ with $n$ punctures. Each Lagrangian description is associated to a pair of pants decomposition of the Riemann surface, 
with the couplings $q_a$ identified with the sewing parameters for the surface. The basic S-dualities of individual $SU(2)$ gauge groups represent basic moves relating different pair of pants decompositions. 

\subsection{Moduli spaces of vacua}

Generically, the moduli space of $\cN=2$ supersymmetric vacua consist of the union of several branches, each factorized into a ``Coulomb branch factor'' and a ``Higgs branch factor''.  
A Coulomb branch factor is parameterized by the vevs $u_i$ of Coulomb branch operators $\cO^i$. 
A Higgs branch factor is an hyper-K\"ahler cone parameterized by the vevs of Higgs branch operators. 
See \cite{Argyres:1998vt} and references therein for more details. 

It is useful to observe that the hyper-K\"ahler geometry of the Higgs factors does not depend on the couplings.
It can thus be usefully computed in convenient corner of parameter space, such as a corner where the theory is weakly coupled. Flavour symmetries act as (tri-holomorphic) isometries on the Higgs branch, and the corresponding mass parameters force the theory to live at fixed points of the corresponding isometries. Often, turning on generic complex masses completely suppresses Higgs branch moduli.

Usually an $\cN=2$ theory has a pure Coulomb branch of vacua, where all Higgs branch operators have zero vet
and $SU(2)_R$ is unbroken. We will refer to this branch simply as the Coulomb branch $\cC$ of the theory. 
At special complex singular loci $\cC_\alpha$ new branches may open up, of the form $\cC_\alpha \times \cH_\alpha$
for some Higgs factors $\cH_\alpha$. 

At low energy on the Coulomb branch, the only massless degrees of freedom are scalar fields which parameterize motion along the Coulomb branch, which sit in Abelian vectormultiplets. 
Thus the low-energy description of physics on the Coulomb branch involves a $U(1)^{r}$ gauge theory, where the rank $r$ is the complex dimension of the Coulomb branch. Supersymmetry implies a close interplay between the couplings of the low energy gauge theory and the geometry of the Coulomb branch. This is the main subject of the next section. 

The supersymmetry algebra in a sector with Abelian (electric,  magnetic and flavour) 
charges $\gamma$ admits a central charge function $Z_\gamma$, linear in $\gamma$ \cite{Witten:1978mh}. 
Schematically, 
\begin{equation}
[Q,Q] = \bar Z \qquad \qquad [Q, \bar Q] = P \qquad \qquad [\bar Q, \bar Q] = \bar Z
\end{equation}
This implies that charged particles are generically massive, with mass above the BPS bound $|Z_\gamma|$ \cite{Prasad:1975kr,Bogomolny:1975de}, and can be integrated out at sufficiently low energy at least at generic points in the Coulomb branch. \footnote{Of course, there could be a separate massless sector which carries no gauge charges. This will happen in theories where the Higgs branch is not fully suppressed at generic points in the Coulomb branch.} 

Particles which saturate the BPS bounds are called BPS particles. They will play an important role in understanding the low energy physics of $\cN=2$ quantum field theories. 

\subsection{Seiberg-Witten theory}
The low energy dynamics in the Coulomb branch is the subject of Seiberg-Witten theory.
The study of the Coulomb branch dynamics was initiated in \cite{Seiberg:1994rs,Seiberg:1994aj}. See also 
e.g. \cite{Lerche:1997sm,Freed:1997dp,ias-volumes} for reviews of the subject.
The central charge function
for a charge vector $\gamma$ including an electric charge $\gamma_e$, a magnetic charge $\gamma_m$ and a flavour charge $\gamma_f$ takes the form 
\begin{equation}
Z_\gamma = a \cdot \gamma_e + a^D \cdot \gamma_m + m \cdot \gamma_f
\end{equation}

The $r$ complex fields $a^I$ are the super partners of gauge fields. 
They give a special local coordinate system, where the metric coincides with the imaginary part $\mathrm{Im} \tau_{IJ}$ of the complexified gauge couplings, which can be packaged locally in an holomorphic prepotential $\cF$:
\begin{equation}
\tau_{IJ} = \frac{\partial^2 \cF}{\partial a^I \partial a^J}
\end{equation}
The dual fields $a^D_I$ are also given in terms of the prepotential 
\begin{equation}
a^D_I = \frac{\partial \cF}{\partial a^I}
\end{equation}
and of course $\tau_{IJ} = \frac{\partial a^D_I}{\partial a^J}$

The prepotential depends generally on the gauge couplings and mass parameters of the theory.
The following relation, valid at fixed masses, is often useful to control the dependence on the couplings:
\begin{equation}
da^I \wedge da^D_I = du^i \wedge dc_i
\end{equation}
Here $u^i$ is the vev of the Coulomb branch operator dual to $c_i$. 

The low energy description is covariant under electric-magnetic dualities. An electric-magnetic duality transformation
rotates the gauge charges by an integer-valued linear transformation which preserves the symplectic pairing
\begin{equation}
\langle \gamma, \gamma' \rangle = \gamma_m \cdot \gamma_e' - \gamma_m' \cdot \gamma_e
\end{equation}
The action on the gauge couplings is simply encoded by an inverse rotation of $(a, a^D)$,
so that the central charge remains invariant. It is also useful to add to the duality group 
redefinitions of the flavour currents by multiples of the gauge currents. These transformations shift $\gamma_f$ by multiples of $\gamma_e$ and $\gamma_m$ and 
correspondingly shift $a$ and $a^D$ by multiples of the mass parameters. We will often denote the set of $(a, a^D)$ as ``periods'', for reasons which will become clear soon. 

The crucial insight of Seiberg and Witten is to realize that there is no electric-magnetic duality frame which is globally 
well-defined over the Coulomb branch. Rather, if we continuously vary the Coulomb branch parameters 
along a closed path which winds around singular loci in the Coulomb branch, we may come back to an electric-magnetic dual description of the original physics. Thus the $(a, a_D)$ are multivalued functions of the Coulomb branch parameters 
$u^i$. It is useful to describe the multi-valuedness in terms of the global structure of the charge lattice $\Gamma$: the charge lattice forms a local system of lattices over the Coulomb branch, with monodromies which preserve the simplectic pairing and the sub lattice $\Gamma_f$ of pure flavour charges. The central charge is a globally defined linear map from $\Gamma$ to the complex numbers.  

The singularities of the Coulomb branch must be loci where additional light degrees of freedom appear. 
In particular, they must be loci where the central charge of some BPS particle goes to zero, as only BPS particles 
can modify the geometry of the Coulomb branch through loop effects. The extra degrees of freedom must assemble themselves into an infrared free or conformal effective description of the low energy theory. If a Higgs branch opens up at the locus, it must be possible to describe it in terms of the low energy degrees of freedom. 

A typical example is a codimension one singularity at which a single BPS hypermultiplet becomes massless. 
Without loss of generality, we can go in a duality frame where the BPS hypermultiplet is electrically charged.
This is an infrared free setup: the BPS hypermultiplet of charge $\gamma$ makes the IR gauge coupling run at one loop as 
\begin{equation}\label{eq:taumono}
\tau_{IJ} \sim - \gamma_I \gamma_J \frac{i}{2\pi} \log a \cdot \gamma
\end{equation}
The behaviour of the magnetic central charges 
\begin{equation}
a^D_I \sim - \gamma_I \frac{i}{2\pi} a\cdot \gamma \log a \cdot \gamma
\end{equation} 
shows the monodromy of the central charge, and thus of the charge lattice:
\begin{equation}
a^D_I \to a^D_I + \gamma_I a\cdot \gamma \qquad q^e_I \to q^e_I - q_m^J \gamma_J \gamma_I
\end{equation}
In a generic duality frame, we can write the monodromy as
\begin{equation} \label{eq:BPSmono}
q \to q - \langle q,\gamma \rangle \gamma
\end{equation}

In general, singular loci where a collection of light, IR free electrically charged BPS hypermultiplets appear will be associated to parabolic monodromies similar to (\ref{eq:BPSmono}). If a sufficiently large number of light particles are present, a Higgs branch may open up, described by the vev of the corresponding hypermultiplet fields. 

Singular loci where the IR description involves a non-trivial superconformal field theory are associated to more general monodromies. 
We expect several periods to go to zero at a superconformal points, scaling as interesting, possibly fractional powers of the Coulomb branch coordinates $u^i$.
Thus the monodromies will be in general elliptic. We are not aware of any example which involves hyperbolic monodromies. 
Interesting superconformal fixed points often arise from the collision/intersection of two or more simple singularities. The collision/intersection of singularities where mutually non-local particles such as an electron and a monopole become light usually produces IR superconformal field theories of the Argyres-Douglas type \cite{Argyres:1995jj,Argyres:1995xn}.

A typical example is the collision of a point where one monopole of charge $1$ is massless, and one point where 
$n_f$ particles of electric charge $1$ become massless. The combined monodromy is
\begin{equation}
a \to a + a_D \to (1-n_f) a + a_D \qquad a_D \to a_D \to -n_f a + a_D
\end{equation} 
and has trace $2 - n_f$. For $n_f=1,2,3$ the monodromy is elliptic, and we obtain an Argyres-Douglas theory which possesses an $SU(n_f)$ flavor symmetry 
rotating the electrically charged particles among themselves.   

The interplay between the monodromies of the charge lattice and the spectrum of BPS particles is rather interesting, and is made more intricate by the phenomenon of wall-crossing. 
General BPS particles belong to supermultiplets which can be described as a (half) hypermultiplet tensored with a spin $j$ representation of the Lorentz group. 
The one-loop contributions from a BPS particle of generic spin $j$ will be proportional to $\Omega_j = (-1)^{2j} (2j+1)$. The sum of $\Omega_j$ 
over all BPS particles with a given charge $\gamma$ is a protected index, which may jump only if the single particle states mix with a continuum of multi particle states
Generically, the mass $|Z_\gamma|$ is larger than the mass of constituents of different charge $|Z_{\gamma'}| + |Z_{\gamma - \gamma'}|$ and the index is protected. 
At walls of marginal stability, where the central charges of particles of different charges align, the index can jump. 
The jumps in the index are controlled by a specific wall-crossing formula due to Kontsevich and Soibelman \cite{ks1}. 

\subsection{Seiberg-Witten curves}
There is a tension between two properties of the matrix of gauge couplings $\tau_{IJ}$: it is locally holomorphic in the Coulomb branch parameters 
$u^i$, and it has a positive-definite imaginary part. In the absence of intricate monodromies, these properties would actually be incompatible with each other. 
There is a rather different mathematical problem where a matrix with very similar properties appear: the period matrix of a family of Riemann surfaces.
Given a Riemann surface, a set of A cycles $\alpha^I$ and dual B cycles $\beta_I$, with 
\begin{equation}
\alpha^I \cap \alpha^J = 0 \qquad \alpha^I \cap \beta_J = \delta^I_J \qquad \beta_I \cap \beta_J =0
\end{equation}
the period matrix $\tau_{IJ}$ is computed from the contour integrals of holomorphic differentials $\omega_I$ on $\beta_J$,
normalized so that the contour integral on $\alpha^J$ is $\delta_I^J$. 

The period matrix has positive definite imaginary part. If we have a holomorphic family of Riemann surfaces, it will depend holomorphically on the parameters, 
with appropriate monodromies around loci where the Riemann surface degenerates. 
Furthermore, if we are given a meromorphic form $\lambda$ on the Riemann surface, such that the variations of $\lambda$ along the family are holomorphic differentials,
the periods of $\lambda$ along $\alpha^I$ and $\beta_I$ will behave in the same way as the periods $a^I$, $a_I^D$. 
More generally, the homology lattice of the Riemann surface behave like the lattice of charges in a gauge theory, with the intersection of cycles 
playing the role of the $\langle, \rangle$ pairing on the charge lattice and the period of $\lambda$ on a cycle $\gamma$ playing the role of 
the central charge $Z_\gamma$. The monodromies around simple degeneration points, where a single cycle $\gamma$
contracts, take exactly the form (\ref{eq:BPSmono}), and the behaviour of the period matrix is precisely (\ref{eq:taumono}).
If $\lambda$ has poles on the Riemann surface, the periods of lambda depend on the homology of the Riemann surface
punctured at the poles, and the residues of $\lambda$ behave like mass parameters. 

Originally, this analogy was used by Seiberg and Witten as a simple computational tool to describe their solution for the low energy dynamics of $SU(2)$ gauge theories with various 
choices of matter. These theories have a one-dimensional Coulomb branch, and the solution was described by simple families of elliptic curves. 
A priori, there was no reason to believe this tool would be useful for theories with a higher-dimensional Coulomb branch: most matrices $\tau_{IJ}$ with positive definite imaginary part 
are not period matrices of a Riemann surface, because the dimension $3g-3$ of moduli space of Riemann surfaces of genus $g$ is much smaller than the dimension of the space of $2g \times 2g$ symmetric matrices. 

Surprisingly, the great majority of known $\cN=2$ field theories do admit a low-energy description in terms of a Seiberg-Witten curve equipped with an appropriate differential. 
For many Lagrangian field theories, this fact can be verified through hard localization calculations (see \cite{Nekrasov:2012xe} for the broadest possible result, and references therein for previous work).\footnote{ It may also be justified through considerations based on surface defects \cite{Gaiotto:2009fs}.}
For class $\cS$ theories, it follows directly from the properties of the six-dimensional SCFTs. 
For theories defined by Calabi-Yau compactifications, the situation is less clear. Almost by construction, the low energy physics can be described by periods of the holomorphic 
three-form on the Calabi-Yau. It is not always obvious if this can be recast in terms of periods of a differential on a Riemann surface.  

We record a useful relation which allows one to associate the UV couplings of Seiberg-Witten theories to 
the corresponding Coulomb branch operators 
\begin{equation}\label{eq:lamlam}
\delta u^i \wedge \delta c_i = \delta a^I \wedge \delta a^D_I = \int_\Sigma \delta \lambda \wedge \delta \lambda
\end{equation}
This is derived through the Riemann bilinear identity.

\subsection{The Coulomb branch of Lagrangian gauge theories}
In a pure $\cN=2$ gauge theory, the D-term equations for the non-Abelian scalar fields $\Phi$ take the form 
\begin{equation}
[ \Phi, \Phi^\dagger ] =0
\end{equation}
Classically, the theory has a family of ($\cN=2$) supersymmetric vacua characterized by a generic complex vev of $\Phi$ belonging to some Cartan sub algebra of the gauge group. If the vev is generic, it Higgses the gauge group down to an Abelian subgroup $U(1)^r$, where $r$ is the rank of the group. The off-diagonal components of the vectormultiplet become massive and can be integrated out at low energy. 

The same analysis typically holds for theories with matter: a generic Coulomb branch vev suppresses the 
vevs of hypermultiplets.\begin{equation}
\Phi \cdot t_{ab} q^b + M \cdot t^f_{ab} q^b =0
\end{equation}

The Coulomb branch survives quantum-mechanically, but the geometry of the Coulomb branch receives important one-loop and instanton corrections. 
At very weak coupling, the electric periods can be identified with eigenvalues $a_I$ of  $\Phi$. The magnetic periods are derived from the 
perturbative pre potential, which only receives tree level and one-loop contributions from the massive W-bosons and hypermultiplets
\begin{equation} \label{eq:lagpre}
\cF = \frac{\tau}{2} a^2 + \sum_{e \in \Delta_+} \frac{i}{2\pi} (a\cdot e)^2 \log a\cdot e- \sum_{(w,w^f) \in R} \frac{i}{4\pi} (a\cdot w + m \cdot w^f)^2 \log (a\cdot w + m \cdot w^f)
\end{equation}
We sum over the positive roots $e$ and the weights for the gauge and flavor representation. 

Seiberg and Witten \cite{Seiberg:1994rs} 
observed that this cannot be the end of the story: because of asymptotic freedom, the coefficient of the logarithms 
makes the gauge couplings negative definite near the locus where a W-boson becomes naively massless.
The pre-potential receives instanton corrections (in the form of a power series in the instanton factors $q$ for the gauge groups)
which must turn the behaviour around, and convert the naive W-boson singularity into  singularities at which the gauge couplings have physically acceptable behaviour.
 
The canonical example is pure $SU(2)$ gauge theory, with Seiberg-Witten curve and differential 
\begin{equation}
x^2 = z^3 + 2 u z^2 + \Lambda^4 z \qquad \lambda = x \frac{dz}{z^2}
\end{equation}
At large values of $u \sim \mathrm{Tr} \Phi^2$ the theory is weakly coupled, and the integral of $\lambda$ on a circle of unit radius in the $z$ plane 
gives $a = \sqrt{2u} + \cdots$. The contour integral along a dual contour gives the expected $a_D = \frac{2i}{\pi} \sqrt{2u} \log u + \cdots$. 
At smaller values of $u$ we encounter two singular loci $u = \pm \Lambda^2$ where a magnetic monopole and a dyon (whose charge add to the W-boson charge $2$)
become respectively massless.

Similarly, the Seiberg-Witten curve for pure $SU(N)$ gauge theory \cite{Klemm:1994qs,Argyres:1994xh} is 
\begin{align}
&y^2 + P_N(x) y + \Lambda^{2N}=0 \qquad \lambda = x \frac{dy}{y}  \cr &P_N(x) = x^N + u_2 x^{N-2} + \cdots + u_N
\end{align}
The naive W-boson singularity at the discriminant of $P(x)$ is replaced by two simple singular loci, at the discriminants of $P\pm \Lambda^N$. 
The self-intersections of the two loci produce interesting Argyres-Douglas singularities. For example, the maximal AD singularity corresponds to the curve 
\begin{align}
&y^2 = x^N + c_2 x^{N-2} + \cdots + u_N \qquad \lambda = x dy 
\end{align}

\section{Theories in the class $\cS$}
The basic starting point for the class $\cS$ construction are the six-dimensional $(2,0)$ SCFTs \cite{Witten:1995zh,Strominger:1996ac,Witten:1995em,Seiberg:1996vs,Seiberg:1997ax,
Seiberg:1997zk}. The known $(2,0)$ SCFTs have an ADE classifications. 
These are strongly-interacting generalizations of the free Abelian $(2,0)$ theory, which consists of a self-dual two-form gauge field, five scalar fields and fermions. The Abelian theory is the world volume theory of a single $M5$ brane. The general SCFTs
arise in M-theory as the world-volume theory of $N$ $M5$ branes (the $A_{N-1}$ theory) \cite{Strominger:1996ac}, possibly in the presence of an $O5$ plane (the $D$-type theories). They also arise in IIB string theory at the locus of an ADE singularity \cite{Witten:1995zh}. The string theory construction of these theories makes two properties manifest.
A SCFT labeled by the Lie algebra $\mathfrak{g}$ 
\begin{itemize}
\item Provides a UV completion to five-dimensional $\cN=2$ SYM theory with gauge algebra $\mathfrak{g}$ 
\item Has a Coulomb branch of vacua where it reduces to an Abelian 6d theory valued in the Cartan of $\mathfrak{g}$, modulo the action of the Weil group. 
\end{itemize}
To be precise, the 6d theory compactified on a circle of radius $R$ should admit an effective description as 5d SYM with gauge coupling $g^2 = R$. 
The two statements are compatible: the 6d Abelian theory on the Coulomb branch compactified on a circle gives a 5d Abelian gauge theory, which also describes the Coulomb branch of 5d SYM. 

Notice that both theories have an $SO(5)_R$ R-symmetry. In the Abelian theories, which are related in the same way, the 
R-symmetry rotates the five scalar fields. The Coulomb branch of the $(2,0)$ SCFT is parameterized by the vevs of Coulomb branch operators, which have the same quantum numbers as the Weil-invariant polynomials in the scalar fields $x^a$ of the Abelian low-energy description \cite{Aharony:1998an,Bhattacharya:2008zy}. The theory on the Coulomb branch has five central charges, which are carried by strings rather than particles. The BPS strings in the theories carry a charge under the Abelian two-form fields, which coincides with a root of $\mathfrak{g}$. The central charges for such a string of charge $e$ are simply $Z^a = e \cdot x^a$. 

The construction of four-dimensional field theories in the class $\cS$ involves a twisted compactification of the SCFTs
on a Riemann surface $C$ \cite{Gaiotto:2009hg}. The twisting uses an $SO(2)$ subgroup of $SO(5)$, and preserves a four-dimensional $\cN=2$ super algebra 
in the four directions orthogonal to the surface. The $SO(2)$ factor becomes $U(1)_r$. The remaining $SO(3)$ becomes $SU(2)_R$. 
The six-dimensional Coulomb branch operators which only 
carry $SO(2)$ charges become Coulomb branch operators for the  $\cN=2$ super-algebra. Notice that due to the twisting
an operator of $SO(2)$ charge $k$ becomes a $k$-form on $C$. The construction of a general theory in the class $\cS$ may involve several further modifications of the theory, which preserve the four-dimensional $\cN=2$ super algebra. We will review some details in a later section. 

These twisted compactifications have a useful property: the Coulomb branch geometry is independent of the area of $C$,
and can be described exactly at large area in terms of vevs of the scalar fields $x$ of $SO(2)$ charge $1$ in the low-energy six-dimensional Abelian description. Because of the twisting, the vevs give a locally holomorphic one-form $\lambda = x dz$ on $C$,
valued in the Cartan of $\mathfrak{g}$ modulo the action of the Weil group.
On the other hand, if we make the area of $C$ is arbitrarily small while keeping the Coulomb branch data fixed, 
we will define a four-dimensional theory which, by definition, is the class $\cS$ theory. 

Thus the Seiberg-Witten low energy description of a class $\cS$ is readily available 
from its definition.  For $A_{N-1}$ theories one can treat $\lambda$ as a single-valued one-form on a Riemann surface $\Sigma$ which is a rank $N$ cover of $C$ \cite{Witten:1997sc}. Then $\Sigma$, $\lambda$ can be identified with the Seiberg-Witten curve and differential for the class $\cS$ theory. Similar approaches work for general $\mathfrak{g}$. 

Much more work is required to find a direct four-dimensional UV descriptions of a given class $\cS$ theory, or to find a class $\cS$ description of a given Lagrangian four-dimensional theory. 
We will first describe the examples involving the $A_1$ theory, where the variety of possible ingredients is more limited, and then sketch the general story. We refer to section 3 of \cite{Gaiotto:2009hg} and to \cite{Gaiotto:2009we}
for a general discussion of the general story unitary theories and \cite{Chacaltana:2012zy} and references therein for a 
more general discussion. 

\subsection{$A_1$ theories}
The twisted $A_1$ 6d theory has a single Coulomb branch operator $\hat \phi_2$ which behaves upon twisting as a quadratic differential on $C$.
The four-dimensional Coulomb branch is thus parameterized by a holomorphic quadratic differential  $\phi_2$, the vev of $\hat \phi_2$. The dimension of the Coulomb branch, for compact $C$ of genus $g$, is $3g-3$. The one-form $\lambda$ satisfies \cite{Gaiotto:2009we,Gaiotto:2009hg}
\begin{equation}\label{eq:A1SW}
\lambda^2 = \phi_2
\end{equation}
This equation defines simultaneously the double-cover $\Sigma$ of $C$ as a curve in $T^*C$, and the differential $\lambda$. 

The complex structure moduli of $C$ are the exactly marginal UV couplings of this class $\cS$ theory. 
There are exactly as many couplings as operators in the Coulomb branch. This can be understood from the 
observation that the Coulomb branch operators which come from $\phi_2$ have the correct $U(1)_r$ charge
to be dual to exactly marginal couplings. We can extract a four-dimensional operator $\hat u^i$
from $\phi_2$ by contracting it a Beltrami differential. It is natural to associate that operator with the 
corresponding complex structure deformation. This can be verified from the relation (\ref{eq:lamlam}) and is discussed in detail in \cite{Gaiotto:2012sf}.

Thus the regions ``at infinity'' of the parameter space of exactly marginal deformations should correspond to 
the boundaries of the complex structure moduli space \cite{Witten:1997sc}, 
where the Riemann surface $C$ degenerates and one or more handles pinch. 
The physical properties of the six-dimensional SCFT confirm this picture. 
Near a degeneration locus we can pick a metric which makes the pinching handle long and thin 
compared to the rest of the surface. In that region, we should be allowed to use the effective description as 5d SYM on a long segment, 
and then find at lower energy a weakly-coupled four-dimensional $SU(2)$ gauge group. 

The 4d gauge coupling 
can be computed and is such that the instanton factor $q$ coincides with the canonical complex structure  parameter which describe the length and twist of the handle. In particular, it becomes weak when the surface pinches. If we go to a maximal degeneration locus, where the Riemann surface reduces to a network of $2g-2$ three-punctured spheres connected by $3g-3$ handles, we will find $3g-3$ $SU(2)$ gauge groups.
The calculation of the periods in this limit agrees with the gauge theory picture. The magnetic periods 
have a logarithmic behaviour which is consistent with the presence of a bloc of trifundamental half-hypermultiplets 
for each three-punctured sphere. As the gauge groups are conformal, we do not expect any other matter fields coupled to the gauge groups. 

This analysis allows us to identify a possible four-dimensional UV description of the class $\cS$ theory associated with a Riemann surface of genus $g$ near a maximal degeneration locus: the $SU(2)^{3g-3}$ theory 
associated to a graph with $g$ loops. The six-dimensional construction provides a global picture of how all 
the S-dual theories are connected through parameter space, and the low-energy Seiberg-Witten description. 

In order to improve our understanding of the physics of decoupling, it is useful to introduce the notion of superconformal defects in the six-dimensional SCFT. A superconformal defect is a local modification of the theory along a hyperplane which preserves the subgroup of the conformal group which fixes the hyperplane, and an appropriate subset of the supercharges. We are interested here in codimension two defects, which preserve a subgroup of the 6d $(2,0)$ superconformal group which is isomorphic to the 4d $\cN=2$ superconformal group. 

Although we have a relatively poor understanding of the six-dimensional theory, there is a simple trick which allows us to define a useful class of defects in terms of the facts we know. We can simply use the twisted compactification strategy to put the theory on a funnel geometry, with an asymptotically flat region connected near the origin to a semi-infinite tube. The configuration preserves $\cN=2$ supersymmetry, and we can flow to the infrared to find something interesting. In the tube region, we flow to the infrared free five-dimensional $SU(2)$ SYM. In the asymptotically flat region, we have the standard 6d theory, modified only at the origin,
in some what which allows it to couple to the 5d SYM theory. Thus construction produces a canonical superconformal defect equipped with an $SU(2)$ flavor symmetry. We will call it the regular defect. 

This construction clarifies what happens in a degeneration limit of $C$: the handle can be removed, leaving behind two regular defects weakly coupled to the corresponding $SU(2)$ gauge group. In general, we can now enrich our starting point, and consider a Riemann surface $C$ of genus $g$ with $n$ regular defects at points of $C$. This gives the six-dimensional realization of the $SU(2)$ quivers associated to a general graph with $g$ loops and $n$ external legs. The use of regular punctures allows us to make contact with standard brane constructions of $\cN=2$ field theories, and verify that the individual three-punctured sphere 
corresponds to a block of trifundamental half-hypermultiplets.

The Seiberg-Witten geometry in the presence of regular defects is still given by (\ref{eq:A1SW}), but the quadratic differential is now allowed a double pole at the location of the punctures: 
\begin{equation}
\phi_2 \sim \left[ \frac{m_a^2}{(z-z_a)^2} + \frac{u_a}{z-z_a} + \cdots \right] dz^2
\end{equation}
Here $m_a$ is the $SU(2)$ mass parameter at the puncture and $u_a$ an extra Coulomb branch parameter dual to the position $z_a$ of the puncture, which is a new exactly marginal coupling. 

Through appropriate decoupling limits, we can go from these four-dimensional SCFTs to more general asymptotically free $SU(2)$ theories, or generalized Argyres-Douglas theories. 
In the six-dimensional description, these examples involve ``irregular'' punctures, where the quadratic differential is allowed poles of order higher than $2$. 
Basic examples are the pure $SU(2)$ Seiberg-Witten theory
\begin{equation}
\phi_2 = \left[ \frac{\Lambda^2}{z} + \frac{2u_a}{z^2} +  \frac{\Lambda^2}{z^3}  \right] dz^2
\end{equation}
and the basic Argyres-Douglas theories 
\begin{equation}
\phi_2 = P_N(z) dz^2
\end{equation}
and 
\begin{equation}
\phi_2 =  \left[ P_N(z) + \frac{u}{z} + \frac{m^2}{z^2} \right] dz^2
\end{equation}
where $P_N(z)$ is a degree $N$ polynomial. 

\subsection{General ADE theories}
The generalization of the $A_1$ results involves several new ingredients. The Coulomb branch is now described by a 
family of differentials associated to the Casimirs of $\mathfrak{g}$, with degree of the differential equal to the degree of the Casimir. 
The exactly marginal couplings still coincide with the space of complex structures of $C$. 
The decoupling limit still replaces a handle by a gauge group with Lie algebra $\mathfrak{g}$,
and can be understood in terms of a codimension $2$ defect with flavor symmetry $\mathfrak{g}$, which we will denote as a full regular defect. 

The first difference is that the theory associated to a three-punctured sphere has a non-trivial Coulomb branch, and no couplings: it is an otherwise unknown
4d SCFT with three $\mathfrak{g}$ flavor symmetries. In order to make contact with standard Lagrangian field theories and their brane engineering we need a larger choice of 
regular punctures. A simple way to understand the possible choices is to realize that the full regular puncture has a Higgs branch, parameterized by the vevs of the moment map operators for the 
 $\mathfrak{g}$ flavor symmetry. The Higgs branch should open up at loci in the Coulomb branch where all the Coulomb branch operators have no pole at the puncture. 
 The Higgs branch conjecturally coincides with the maximal complex nilpotent orbit of $\mathfrak{g}^{\mathbb{C}}$. 
 
 If we sit at a generic point of the Higgs branch and flow to the IR, we essentially erase the puncture. If we sit at a non-generic nilpotent element and flow to the IR, we will 
 somewhat ``simplify'' the full regular puncture to a different type of regular puncture, where the singularities of the Coulomb branch operators are 
 constrained in appropriate patterns. A nilpotent element can always be taken to be the raising operator of an $\mathfrak{su}(2)$ subalgebra $\rho$ of $\mathfrak{g}$.
 Thus these new regular punctures will be labelled by $\rho$. A further generalization of regular punctures is possible, in which the operators of the 6d theory 
undergo a monodromy around the defect, under an outer automorphism of $\mathfrak{g}$ \cite{Chacaltana:2012zy}. 
These general regular punctures allow one to make contact with most superconformal Lagrangian quiver gauge theories. 
As for the $A_1$ case, one can also define a large variety of irregular punctures. 

\section{Calabi-Yau compactifications}
The compactification of string theory on non-compact Calabi-Yau manifolds can also give rise to four-dimensional $\cN=2$ field theories
\cite{Katz:1996fh}. 
The low-energy dynamics can be derived in a straightforward way from the geometry in a type IIB duality frame:
the periods are identified with the periods of the holomorphic three-form on appropriate cycles in the geometry. 
On the other hand, the identification of an intermediate UV-complete four-dimensional field theory description 
of the theory is more laborious. Often, the field theory is engineered through a type IIA construction, and then 
mirror symmetry gives the map to IIB string theory and thus the low energy solution of the theory. 

Theories in the class $\cS$ can be embedded in Type IIB string theory by engineering the 6d SCFTs as loci of ADE singularities, fibered appropriately over the 
curve $C$. For example, an $A_1$ theory can be realized through the geometry 
\begin{equation}
x^2 + u^2 + v^2 = \phi_2(z)
\end{equation}

The geometric engineering, though, can provide solutions for theories which do not admit a known six-dimensional construction, such as 
quiver gauge theories in the shape of E-type Dynkin diagrams. Indeed, it provides a unified picture of all the quiver gauge theories of unitary groups, 
through geometries where an elliptic singularity is fibered over a complex plane \cite{Katz:1997eq}. Remarkably, this provides a description of the space of exactly marginal deformations 
as a moduli space of flat connections on a torus. 

A second remarkable example is a large family of Argyres-Douglas theories, labeled by two ADE labels. 
Remember the $A_1$ examples, lifted to a Calabi-Yau
\begin{equation}
u^2 + v^2 + x^2 + P_N(z) =0
\end{equation}
The $A_{M-1}$ generalization is 
\begin{equation}
u^2 + v^2 + x^M + z^N + \cdots =0
\end{equation}
The main idea is to write that in terms of ADE polynomials for $A_{N-1}$ and $A_{M-1}$ 
as 
\begin{equation}
W_{A_{N-1}}(u,x) + W_{A_{M-1}}(v,z)=0
\end{equation}
and then replace either polynomials with the ones associated to $D$ type, $u^2 x + x^N$, or $E$ type 
\begin{equation}
u^3 + x^4 \qquad u^3 + u x^3 \qquad u^3 + x^5
\end{equation}
 
 \newpage
 \providecommand{\href}[2]{#2}\begingroup\raggedright
\paragraph{\large References to articles in this volume}
\renewcommand{\refname}{

\paragraph{Other references}
\renewcommand{\refname}{\endgroup
\end{document}